\documentclass[aps,prd,preprint,showpacs,showkeys,eqsecnum,nofootinbib]{revtex4-1}

\usepackage{epsfig}
\usepackage{amsmath}
\usepackage{amssymb}
\usepackage{amsfonts}
\usepackage{natbib}

\begin{document}

\title{\bf Diffractive Higgs boson photoproduction in Ultraperipheral Collisions at LHC}

\author{M. B. Gay Ducati}
\email{beatriz.gay@ufrgs.br}

\author{G. G. Silveira}
\email{gustavo.silveira@ufrgs.br}

\affiliation{High Energy Physics Phenomenology Group, Instituto de F\'{i}sica, Universidade Federal do Rio Grande do Sul, Caixa Postal 15051, CEP 91501-970 - Porto Alegre, RS, Brazil.}

\begin{abstract}
A new production mechanism for the Standard Model Higgs boson in Ultraperipheral Collisions at LHC, which allows Central Exclusive Diffractive production by Double Pomeron Exchange in photon-proton processes, is presented. The Higgs boson is centrally produced by gluon fusion with two Large Rapidity Gaps emerging in the final state, being the main experimental signature for this process. As already studied for Pomeron-Pomeron and two-photon processes, the Higgs boson photoproduction is studied within this new mechanism in proton-proton ($pp$) and proton-nucleus ($pA$) collisions, where each system has a different dynamics to be taken into account. As a result, this mechanism predicts a production cross section for $pp$ collisions of about 1.8 fb, which is similar to that obtained in Pomeron-Pomeron processes. Besides, in $pPb$ collisions the cross sections have increased to about 0.6 pb, being comparable with the results of two-photon processes in $pAu$ collisions. Therefore, as the Rapidity Gap Survival Probability is an open question in high-energy Physics, an analysis for different values of this probability shows how competitive are the mechanisms in the LHC kinematical regime.
\end{abstract}



\maketitle

\section{INTRODUCTION}\label{sec:intro}

Different processes have been studied to investigate the Higgs boson production at LHC \cite{Carena:2002es,Hahn:2006my} in order to improve its signal-to-background ($S/B$) ratio. Moreover, experimental results had been used to find out the mass ranges where there is a higher probability to detect the Higgs boson. The analysis of the LEP data set the lower bound of $M_{H} \geq$ 114.4 GeV for the Higgs boson mass \cite{Barate:2003sz} and, on the other hand, the analysis of the new data collected by the CDF and D0 experiments in the Tevatron have excluded the range 160 GeV $< M_{H} <$ 170 GeV \cite{:2009pt,Collaboration:2009je}. Then, it is available for the detection of the Higgs boson the mass range from 115 GeV to 160 GeV, which is a favorable region to observe its decay into bottom quarks if $M_{H} < 135$ GeV, or into electroweak vector bosons for $M_{H} > 135$ GeV \cite{Plehn:2009nd}. However, only the decays into $\gamma\gamma$ and $\tau^{+}\tau^{-}$ are measurable in LHC \cite{Carminati:2004fp,D'Enterria:2007xr}.

The study of diffractive processes may be a richer possibility to significantly increase the $S/B$ ratio \cite{Khoze:2001xm}, since these processes provide a less populated final state and characteristic signatures. Regarding the Regge theory in diffractive processes, different mechanisms for the Central Exclusive Diffractive (CED)  Higgs boson production were studied in the past \cite{Bialas:1991wj,Khoze:1997dr}. The Double Pomeron Exchange (DPE) has an additional gluon exchange in the $t$-channel, if compared to the direct process, which neutralizes the color flow during the interaction. Thus, in this mechanism, the Durham group predicts a production cross section of 3 fb \cite{Khoze:2007td}, being the Large Rapidity Gaps (LRG) the main signature for its detection. Apart from Pomeron exchanges, the Higgs boson may be explored in electroweak interactions in two-photon processes, which are naturally a diffractive process. Even gathering more contributions to the production vertex \cite{Miller:2007pc}, the two-photon mechanism predicts a cross section in $pp$ collisions of about 0.1 fb \cite{Levin:2008gi}, which is less promising in $pp$ collisions than the Pomeron-Pomeron process.

The predictions for the two-photon processes are enhanced in $pA$ and $AA$ collisions, where the strong electromagnetic flux is increased due to the large number of charged particles into the nuclei. This approach was formulated through the Equivalent Photon Approximation (EPA) \cite{Muller:1990wc,Cahn:1990jk}, allowing to factorize the hadronic cross section by the convolution of the $\gamma p$ cross section with the photon flux from the source object. As a result, the photon flux in Ultraperipheral Collisions (UPC) is obtained from the electromagnetic dynamics of charged particles in ultrarelativistic collisions \cite{Budnev:1974de}, such that the cross section for the Higgs boson production is predicted as 0.64 pb for $pAu$ collisions, and 3.9 nb for $AuAu$ collisions \cite{Levin:2008gi}, which are substantially higher than the predictions of the Pomeron-Pomeron mechanism, those being 0.1 pb and 3.92 pb, respectively. Indeed, there are some projects to perform $pA$ and $AA$ collisions in LHC after the runs with protons \cite{Carminati:2004fp,D'Enterria:2007xr}, scheduled for this year \cite{D'Enterria:priv}. Some nucleus species are being considered to compose the particle beams, and this choice will affect the particle dynamics for specific processes, like the beam luminosity for the Higgs boson production. For electromagnetic processes, this property is an advantage for $pA$ collisions if compared to $pp$ and $AA$ ones \cite{d'Enterria:2009er}, which plays an important role for the Higgs boson production in the two-photon mechanism.

Therefore, we investigate a new production mechanism for the CED Higgs boson through the $\gamma p$ subprocess in UPC, which includes some details from both two-photon and Pomeron-Pomeron approaches. For this purpose, the cross section is evaluated for hadronic collisions in LHC, and we investigate the $pA$ collisions in the LHC kinematical regime as means to stress the advantages of the proposed mechanism. This paper is organized as follows: in Section \ref{sec:amp} the $\gamma p$ cross section is presented, with the calculation based on the Impact Factor Formalism (IFF) \cite{GayDucati:2008zs}. Next, in Section \ref{sec:had} this approach is applied to UPC through the EPA in order to compute the production cross section. In Section \ref{sec:res} the results of the photoproduction approach are presented for both $pp$ and $pA$ collisions. After that, Section \ref{sec:disc} discusses an important aspect of the proposed mechanisms and its relevance over other approaches. As a matter of fact, the Rapidity Gap Survival Probability (GSP) is an important phenomenological parameter to be considered in order to get reliable predictions for the Higgs boson production at LHC. Finally, Section \ref{sec:ccl} summarizes the conclusions of this work.

\section{PHOTON-PROTON SUBPROCESS}\label{sec:amp}

The $\gamma p$ subprocess was already explored in the Tevatron and LHC kinematical regimes, computing the event rate with a Sudakov form factor in Double Logarithm Accuracy (DLA) and a GSP of 3\% for LHC and 5\% for the Tevatron \cite{GayDucati:2008zs}. In order to increase the production cross section in $pp$ collisions, we present the production mechanism of the Higgs boson in UPC, including the Sudakov form factor in Leading Logarithm Accuracy (LLA), a wider range for the GSP in LHC, and more contributions for the gluon PDF.

\begin{figure}[t]
\includegraphics*[scale = 0.7]{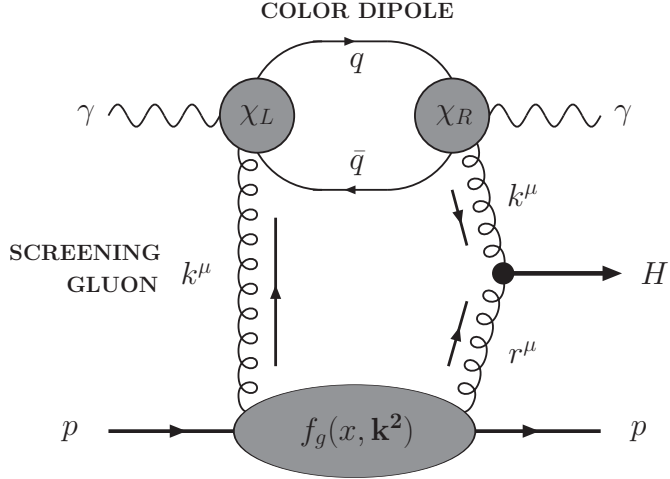}
\caption{\label{fig1}Feynman diagram showing the production mechanism of the Higgs boson in $\gamma p$ processes. $\chi_{L}$ and $\chi_{R}$ are the effective vertices for the coupling of the photon to the $t$-channel gluon, computing the scattering amplitude in the IFF. The configuration of the gluon lines expresses the interaction by DPE, and the large Higgs boson vertex includes the triangle loop with the leading contribution of the top quark. Considering that the proton transfers momentum during the interaction, $f_{g}(x,\boldsymbol{k}^{2})$ is the off-diagonal unintegrated gluon distribution function.}
\end{figure}

The process related to this production mechanism is shown in Fig.~\ref{fig1}. The initial photon splits into a $q\bar{q}$ pair (or color dipole) with which the proton interacts by DPE. Once the gluons are exchanged in the $t$-channel, the $q$ and $\bar{q}$ recombine into a real photon, providing the $\gamma H$ final state with LRG. In order to compute the scattering amplitude of this process, we introduce the IFF, where the Feynman rules can be applied in a straightforward way.

The scattering amplitude for the $\gamma p$ subprocess with the CED Higgs boson production is given by \cite{GayDucati:2008zs}
\begin{eqnarray}
(\textrm{Im}{\cal{A}})_{T} = - \frac{s}{6} \frac{M^{2}_{H}}{\pi v} \frac{\alpha_{s}}{N_{c}} \, \left( \frac{\alpha_{s} C_{F}}{\pi} \right) \int \frac{1}{\boldsymbol{k}^{2}\boldsymbol{k}^{2}\boldsymbol{r}^{2}} \; \Phi^{T}_{\gamma\gamma}(\boldsymbol{k}^{2},Q^{2}) \; d\boldsymbol{k}^{2},
\label{amp-im-1}
\end{eqnarray}
where $v$ = 246 GeV is the vacuum expectation value of the Electroweak Theory, and $\Phi^{T}_{\gamma\gamma}$ is the transverse polarization impact factor of the $q\bar{q}$ fluctuation \cite{Forshaw:1997dc}
\begin{eqnarray}
\Phi^{T}_{\gamma\gamma} (\boldsymbol{k}^{2},Q^{2}) = 4 \pi \alpha_{s} \alpha \sum_{q}^{n_{f}} e^{2}_{q} \int_{0}^{1} \frac{\boldsymbol{k}^{2}[\tau^{2} + (1 - \tau)^{2}][\alpha^{2} + (1 - \alpha)^{2}]}{\boldsymbol{k}^{2} \tau (1 - \tau) + Q^{2} \alpha(1 - \alpha)} \; d\alpha \, d\tau .
\label{impt-fact-gg}
\end{eqnarray}
Considering the forward scattering limit, one may approximate $(\boldsymbol{k}^{2}\boldsymbol{k}^{2}\boldsymbol{r}^{2})^{-1} \to \boldsymbol{k}^{-6}$, which is a particular result of $\gamma p$ process \cite{GayDucati:2008zs}.

Until now this scattering amplitude was evaluated at partonic level. So that, to account for the amplitude of the $\gamma p$ process one has to replace the contribution of the $qg$ coupling to the proton one, which means to take a Partonic Distribution Function (PDF) into account. This replacement introduces the Pomeron coupling to the proton as a gluon ladder, i.e., introduces an off-diagonal unintegrated PDF $f_{g}(x,\boldsymbol{k}^{2})$ \cite{Khoze:1997dr}
\begin{eqnarray}
\frac{\alpha_{s}C_{F}}{\pi} \to f_{g}(x,\boldsymbol{k}^{2}) = {\cal{K}} \left( \frac{\partial [xg(x,\boldsymbol{k}^{2})]}{\partial \ln \boldsymbol{k}^{2}} \right),
\label{repl-pom_proton}
\end{eqnarray}
where $xg(x,\boldsymbol{k}^{2})$ is the integrated gluon distribution function, and ${\cal{K}} = (1.2)\exp(-B\boldsymbol{p}^{2}/2)$ is a multiplicative factor to take into account the off-diagonality of the distribution \cite{Shuvaev:1999ce}, with $B$ = 5.5 GeV$^{-2}$. As the imaginary part of the amplitude is the leading contribution to this process, one integrates the squared amplitude [Eq.(\ref{amp-im-1})] to get the event rate for the CED Higgs boson production in $\gamma p$ processes, which reads
\begin{eqnarray}
\frac{d\sigma}{d \boldsymbol{q}^{2} dy_{_{H}}} = \frac{\alpha_{s}^{4} \, K_{NLO}}{288 \pi^{5} b } \left( \frac{M^{2}_{H}}{N_{c}v} \right)^{2} \left[ \int \frac{d\boldsymbol{k}^{2}}{\boldsymbol{k}^{6}} \; f_{g}(x,\boldsymbol{k}^{2}) \; \Phi^{T}_{\gamma\gamma}(\boldsymbol{k}^{2},Q^{2}) \right]^{2},
\label{final-eq1}
\end{eqnarray}
with $K_{NLO}$=1.5 being the $K$-factor related to the NLO contributions to the $ggH$ vertex \cite{Spira:1995rr}.

This theoretical result expresses the use of the Feynman rules for the process presented in Fig.~\ref{fig1}, however one should mind about the physics related to this particular process, which is going to bring up some important corrections. As the gluon momentum $k$ goes to zero, many gluons will be emitted from the production vertex if the screening gluon does not neutralize the color flow during the interaction. Even so, it is necessary to account for the contributions from virtual QCD diagrams that bring out terms like ln($M_{H}^{2}/\boldsymbol{k}^{2}$). The emission probability of one gluon in Double Logarithm Approximation (DLA) is given by
\begin{eqnarray}
S(\boldsymbol{k}^{2},\mu^{2}) = \frac{ N_{c}\alpha_{s} }{\pi} \int^{\mu^{2}}_{\boldsymbol{k}^{2}} \frac{d\boldsymbol{p}^{2}}{\boldsymbol{p}^{2}} \int^{M_{H}/2}_{|\boldsymbol{p}|} \frac{dE}{E} = \frac{3N_{c} }{ 4\pi } \; \ln^{2} \left( \frac{ M^{2}_{H} }{4\boldsymbol{k}^{2}} \right),
\label{s-sudakov}
\end{eqnarray}
where $\mu = M_{H}/2$, and $E$ and $p$ are the energy and momentum of the emitted gluon, respectively. The lower limit in the first integral corresponds to the lower momentum allowed for a gluon emission in the $s$-channel. The suppression of several gluon emissions exponentiates, and then the non-emission probability has the form $\exp[-S(\boldsymbol{k}^{2},\mu^{2})]$. This factor has to be included to the event rate in order to guarantee that the integral over the gluon momentum is regulated in the infrared region \cite{Bartels:2006ea}. In other words, it means that we account for the events where the neutralization during the interaction has to be effective.

In addition to the gluon emissions, other contributions come from the possibility of a quark being emitted from the production vertex, and these diagrams lead to the emission probability in Leading Logarithm Approximation (LLA). Thus, the emission probability in LLA includes both gluon and quark emission, changing Eq.(\ref{s-sudakov}) to
\begin{eqnarray}
T(\boldsymbol{k}^{2},\mu^{2}) = \int^{\mu^{2}}_{\boldsymbol{k}^{2}} \frac{\alpha_{s}(\boldsymbol{p}^{2})}{2\pi} \frac{d\boldsymbol{p}^{2}}{\boldsymbol{p}^{2}} \int^{1-\Delta}_{0} \left[ zP_{gg}(z) + \sum_{q} P_{qg} (z) \right] dz,
\label{s-sudakov-lla}
\end{eqnarray}
where $\Delta = 2|\boldsymbol{p}|/M_{H}$, and $P_{gg}$ and $P_{qg}$ are the LO DGLAP splitting functions. There are recent results showing that the scale could be taken as $\mu = M_{H}$ \cite{Coughlin:2009tr}, however we will keep in this work the prescription $\mu = M_{H}/2$ \cite{Khoze:2007hx}, for coherence with our  previous work \cite{GayDucati:2008zs}.

Solving Eq.(\ref{s-sudakov-lla}) leads to the following expression for the form factor
\begin{eqnarray}
T(\boldsymbol{k}^{2},\mu^{2}) = \left[ \frac{\alpha_{s}(\boldsymbol{k}^{2})}{\alpha_{s}(\mu^{2})} \right] e^{-S(\boldsymbol{k}^{2},\mu^{2})} .
\label{t-sudakov-final}
\end{eqnarray}
which is a more accurate result for the probability of bremsstrahlung gluons, since this probability is overestimated in DLA \cite{Khoze:2000cy}. In this case, the strong coupling constant is running instead of the one in Eq.(\ref{final-eq1}), which is fixed to 0.2. Proceeding with the summation of all leading logarithms that contribute to the scattering amplitude, one has to include Eq.(\ref{t-sudakov-final}) to the unintegrated gluon PDF in order to incorporate all contributions from virtual diagrams \cite{Dokshitzer:1978hw}
\begin{eqnarray}
\tilde{f}_{g}(x,\boldsymbol{k}^{2},\mu^{2}) = {\cal{K}} \, \frac{\partial}{\partial \ln \boldsymbol{k}^{2}} \left[ \sqrt{T(\boldsymbol{k}^{2},\mu^{2})} \, xg(x,\boldsymbol{k}^{2}) \right],
\label{sud-t-pdf}
\end{eqnarray}
where the square root denotes that the contribution is important for the hard gluon, since for $x^{\prime} \ll x$ only the self-energy of the hard gluon contributes in LLA \cite{Martin:1997wy}. As one could see from Eq.(\ref{final-eq1}), the integrand is divergent as $\boldsymbol{k}$ goes to zero. The inclusion of this form factors is important to regulate the scattering amplitude in the infrared region, vanishing the non-emission probability faster than $\boldsymbol{k}^{-6}$ \cite{Forshaw:2005qp}.

Finally, including all corrections, the event rate is given by
\begin{eqnarray}
\frac{d\sigma}{d \boldsymbol{q}^{2} dy_{_{H}}} = S^{2}_{gap} \frac{\alpha_{s}^{4} \, K_{NLO}}{288 \pi^{5} b } \left( \frac{M^{2}_{H}}{N_{c}v} \right)^{2} \left[ \int_{\boldsymbol{k}^{2}_{0}}^{\mu^{2}} \frac{d\boldsymbol{k}^{2}}{\boldsymbol{k}^{6}} \; \tilde{f}_{g}(x,\boldsymbol{k}^{2},\mu^{2}) \; \Phi^{T}_{\gamma\gamma}(\boldsymbol{k}^{2},Q^{2}) \right]^{2}.
\label{final-eq1-sud}
\end{eqnarray}
An important aspect related to the integral over $\boldsymbol{k}$ is the lower limit $\boldsymbol{k}^{2}_{0} = 0.3$ GeV$^{2}$, which is added to extend the gluon distribution function to values of $\boldsymbol{k}^{2}$ lower than the initial scale of evolution used by the parametrizations for the gluon PDF \cite{Martin:2002dr,Martin:2009iq,Martin:2009bu}. The distribution function in this region is parametrized with a particular function of $\boldsymbol{k}^{2}$ that goes to zero in the limit $\boldsymbol{k}^{2} \to 0$. Thus, when we apply each parametrization, we need to adjust the parameters to fit the distribution in both regions, taking the transition point $\boldsymbol{k}^{2}$ = 1.25 GeV$^{2}$.

Furthermore, to properly predict the number of diffractive events that is going to be observed in LHC, the Gap Survival Probability $S^{2}_{gap}$ was added to Eq.(\ref{final-eq1-sud}), for which we assume two distinct values: (i) in comparison to the values predicted in a previous work, we consider a GSP of 3\% for LHC, and (ii) a probability of 10\% based on the ratio for dijets production in HERA \cite{Kaidalov:2003xf}. There are some important remarks concerning this aspects that will be discussed in Section \ref{sec:disc}.

\section{HADRONIC CROSS SECTION}\label{sec:had}

For collisions of charged particles with large impact parameters, strong interactions can be neglected by impact parameter cut \cite{Baltz:2007kq}. The photon flux from a relativistic charged source is so intense that its contribution to the photoproduction cross section is significative. Then, one may compute the cross section taking into account the possibility of particle production due to the electromagnetic emissions from the colliding particles. The hadronic cross section for the Higgs boson photoproduction in the collision of fast moving charge particles is factorized as
\begin{eqnarray}
\sigma_{had} = 2 \int_{\omega_{min}}^{\omega_{max}} \frac{dn}{d\omega} \, \sigma_{\gamma p}(\omega) \, d\omega
\label{xsec-had-upc}
\end{eqnarray}
where $\omega_{min} = M^{2}_{H}/2x\sqrt{s_{NN}}$ is the minimum photon energy to produce the Higgs boson, with $\sqrt{s_{NN}}$ being the center-of-mass energy of the $pA$ system, which is defined in Tab.~\ref{tab1}, and $x$ the momentum fraction of the proton carried by the gluon. The photon flux $dn_{i}/d\omega$ is exponentially suppressed in the high-energy limit, such that $\omega_{sup} \approx \gamma_{L}/R$ defines the energy from which the flux is suppressed \cite{Baltz:2007kq}. The convolution of the photon flux with the $\gamma p$ cross section is performed on the dependence of the photon energy. In the $\gamma p$ cross section, the photon  energy dependence appears in the decomposition of the photon virtuality into the Sudakov parametrization \cite{Motyka:2008ac}
\begin{eqnarray}
q^{2} = -Q^{2} = - \frac{\omega^{2}}{\gamma^{2}_{L}\beta^{2}_{L}} - \boldsymbol{q}^{2}.
\label{Q2-decomp}
\end{eqnarray}
Thus, the upper limit $\omega_{max} = \sqrt{Q^{2}\gamma^{2}_{L}\beta^{2}_{L}}$ in Eq.(\ref{xsec-had-upc}) determines the maximum energy of the emitted photon. In UPC, the photon virtuality is related to the size of the source object, and has an upper bound defined by the coherent condition for photon emissions, such that
\begin{eqnarray}
Q^{2} \lesssim \frac{1}{R^{2}},
\end{eqnarray}
where for protons this upper bound is 0.04 GeV$^{2}$ \cite{GayDucati:2008zs}. At this point, $s_{NN}$ is the center-of-mass energy squared of the $pp$ or $pA$ system, and $W^{2} = 2\omega\sqrt{s_{NN}}$ denotes the center-of-mass energy squared of the $\gamma p$ subprocess. This subsystem is not symmetric in the interchange of the photon source and the target, such that other contribution arises by the replacement $y_{_{H}} \to - y_{_{H}}$, affecting the momentum fraction of the proton $x = (M^{2}_{H}/W^{2}) \, e^{\pm y_{_{H}}}$ carried by the gluon. Then, the total cross section is given by the sum of this contributions
\begin{eqnarray}
\sigma_{tot} = 2 \int_{\omega_{min}}^{\omega_{max}} \frac{dn_{i}}{d\omega} \, \sigma_{\gamma p}(\omega) \, d\omega + (y_{H} \to - y_{H})
\label{xsec-had-total}
\end{eqnarray}

The photon flux $dn_{i}/d\omega$ has a particular expression for each source object, which could be a proton or a nucleus. When a proton is taken as the source object, the photon flux is given by \cite{Klein:2003vd}
\begin{eqnarray}
\frac{dn_{p}}{d\omega} = \frac{\alpha_{em}}{2\pi\omega} \left[ 1 + \left( 1 - \frac{2\omega}{\sqrt{s_{NN}}} \right)^{2} \right] \left( \mbox{ln}\mu_{p} - \frac{11}{6} + \frac{3}{\mu_{p}} - \frac{3}{2\mu_{p}^{2}} + \frac{1}{3\mu_{p}^{2}} \right),
\label{eq-flux-p}
\end{eqnarray}
with $\mu_{p} = 1 + (0.71\mbox{ GeV}^{-2})\sqrt{s}/2\omega^{2}$. However, for the case of a nucleus as the source object, the photon flux has a different form due to the nuclear density. This distribution can be found analytically as \cite{Klein:1999qj}
\begin{eqnarray}
\frac{dn_{A}}{d\omega} = \frac{2Z^{2}\alpha_{em}}{\pi\omega} \left\{ \mu_{A} K_{0}(\mu_{A}) K_{1}(\mu_{A}) - \frac{\mu_{A}^{2}}{2} \left[ K^{2}_{1}(\mu_{A}) - K^{2}_{0}(\mu_{A}) \right] \right\},
\label{eq-flux-A}
\end{eqnarray}
where $\mu_{A} = \omega b_{min} / \gamma_{L}$, with $b_{min} = r_{p} + R_{A}$. The factor $\gamma_{L} = (1 - \beta_{L}^{2})^{-1/2} = \sqrt{s_{NN}}/2m_{i}$ is the Lorentz factor of a single beam, with $m_{p}$ = 0.938 GeV the proton mass, and $m_{N}$ = 0.9315 GeV the nucleon mass \cite{Baltz:2007kq}. Tab. \ref{tab1} shows all kinematical parameters introduced in this calculation. In the case of $pA$ collisions in the photoproduction mechanism, there are the possibilities of (i) the photon being emitted from the nucleus, and interacting with the proton, or (ii) the proton emits a photon, which will interact with the nucleus. For the Higgs boson photoproduction, we just consider the contribution from the first possibility, since in the second case a nuclear PDF would decrease the cross section in the region $x < 0.01$.
\begin{table}[t]
\begin{tabular}{||c|c|c|c|c|c||}\hline\hline
System & $\sqrt{s_{NN}}$ & $R$  & $\omega_{max}$ & $\gamma_{L}$ \\
       & (TeV)           & (fm) & (GeV)          &              \\ \hline
$pp$   & 14              & 0.7  & 2102           & 7460         \\
$pO$   & 9.90            & 3.0  & 350            & 5314         \\
$pAr$  & 9.40            & 4.1  & 256            & 5045         \\
$pAu$  & 8.86            & 7.0  & 135            & 4755         \\
$pPb$  & 8.80            & 7.1  & 133            & 4724         \\ \hline\hline
\end{tabular}
\caption{\label{tab1}Input parameters to compute the photoproduction cross section: (i) center-of-mass energy given by $\sqrt{s_{NN}}$ = (14 TeV) $\sqrt{Z_{A}Z_{B}/AB}$, (ii) radius $R_{A} = r_{0}A^{1/3}$ of the source object that emits the real photons, with $r_{0}$ = 1.2 fm, (iii) the photon energy $\omega_{sup} \approx \gamma_{L}/R_{A}$ from where the photon flux is being suppressed, and (iv) the Lorentz factor $\gamma_{L} = \sqrt{s_{NN}}/2m_{i}$ of a single beam, with $m_{p}$ = 0.9383 GeV for protons, and $m_{N}$ = 0.9315 GeV for nucleons.}
\end{table}

From the point of view of high-energy phenomenology, the cross section for the electromagnetic production of the Higgs boson is enhanced if one takes a nuclei as the source object, since the photon flux is enhanced by a factor of $Z^{2}$ for $pA$ and $Z^{4}$ for $AA$ collisions. However, this scenario is modified if one considers the experimental aspects of the beam dynamics in LHC. There are several advantages of $pA$ collisions over the $pp$ and $AA$ ones in LHC, which lead us to consider the former to make the predictions for the photoproduction mechanism. For $pA$ collisions, the luminosity is about four orders of magnitude lower than the $pp$ luminosity, but considering the pile-up and the improvement by $Z^{2}$ in the cross section, the $pA$ collisions arise as a promising way to study the production of high-mass systems \cite{d'Enterria:2009er}. For completeness, due to the low luminosity, especially for particle beams composed by nucleus with high $Z$, the $AA$ collisions do not encourage the study of the Higgs boson production in LHC. Apart of the beam dynamics, the photoproduction mechanism has an additional  advantage for $pA$ collisions instead of the $AA$ ones, since, in the latter, one would take into account a nuclear PDF, which decreases substantially the cross section due to the shadowing effect.

\section{RESULTS}\label{sec:res}

\begin{figure}[t]
\includegraphics*[scale = 0.5]{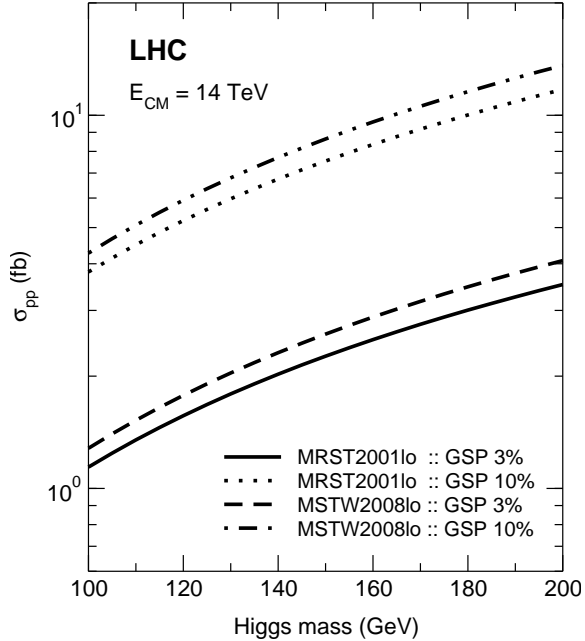}
\caption{\label{fig2}The production cross section for the SM Higgs boson in $pp$ collisions at LHC. The distinct curves present two different possibilities for the GSP in the photoproduction mechanism, as well as different parametrizations for the gluon PDF.}
\end{figure}

Our first goal in this work is to compute the photoproduction cross section of the Higgs boson in $pp$ collisions at LHC, and to compare these results with those obtained in other production mechanisms \cite{Khoze:2007td,Levin:2008gi,d'Enterria:2009er}. Employing more phenomenological corrections to the production cross section as has been done in our previous work, it has enhanced our predictions in one order of magnitude. The Fig.~\ref{fig2} presents the results for the photoproduction cross section for $pp$ collisions in the mass range expected to observe the Higgs boson. The distinct curves show the results using the two possibilities for the GSP in this mechanism. Moreover, these results are evaluated with two distinct parametrizations for the gluon PDF in order to be compared with those employed in Ref.\cite{Forshaw:2005qp} with the MRST2002LO parametrization \cite{Martin:2002dr}, but also with the MSTW2008LO parametrization \cite{Martin:2009iq,Martin:2009bu}. Clearly, the results present an improvement with the use of the MSTW2008LO parametrization. For a GSP of 3\%, the photoproduction cross section is about 2 fb, which is one order higher than the results from the two-photon mechanism. However, with a GSP of 10\% the cross section increases to about 6 fb, being almost twice the cross section predicted from the Pomeron-Pomeron mechanism. Then, combining the DPE to centrally produce the Higgs boson in $\gamma p$ processes and the photon emission from a relativistic colliding particle appears as a promising way to look for the Higgs boson detection in LHC.

\begin{figure}[t]
\includegraphics*[scale = 0.5]{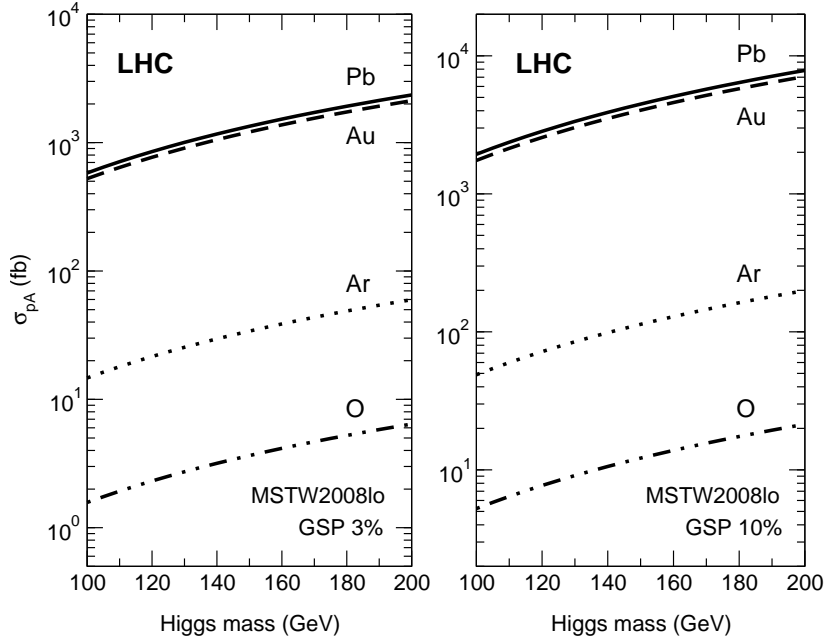}
\caption{\label{fig3}Results for the production cross section for the SM Higgs boson production in LHC considering future projects for $pA$ collisions. The results were evaluated considering the MSTW2008LO parametrization for the gluon PDF.}
\end{figure}

Furthermore, in Fig.~\ref{fig3} we present the results to explore the LHC kinematical regime for $pA$ collisions in order to investigate the importance of such collisions for the Higgs boson production. Some nucleus species are being considered to compose the particles beams in future projects of LHC \cite{Carminati:2004fp,D'Enterria:2007xr}, such that recent works carried out predictions regarding some of them for the Higgs boson production by the two-photon processes \cite{d'Enterria:2009er,Levin:2008gi}. In the case of the photoproduction mechanism, we evaluated the cross sections for $Pb$ and $Au$, which have a higher $Z$, but also for species with smaller $Z$, like $O$ and $Ar$. Although the $Au$ nucleus has not been considered to compose the particle beams in LHC, we inspect the cross section to compare with previous results of two-photon processes. All species investigated in this work are presented in Tab. \ref{tab1} with the respective parameters considered in this calculation.

The results for the photoproduction cross section are computed for $pA$ collisions, inspecting it for the nuclei species presented in Tab. \ref{tab1}. As the species have very different $Z$, the cross sections enhance significantly for heavier nuclei, which increases the event rate for the Higgs boson production. In this case, we employed the MSTW2008LO parametrization for the gluon PDF into the proton, and used the same values for the GSP as before. Comparing these results with those for $pp$ collisions, the cross section is enhanced by a factor higher than 10$^{2}$ for the $pPb$ nucleus, which is similar to that obtained in the two-photon process in Ref.~\cite{Levin:2008gi} for the $Au$ nucleus. On the other hand, both these results are significantly higher in comparison to the results in Ref.~\cite{d'Enterria:2009er}, which predict a production cross section of 170 fb. Therefore, nuclei collisions are very worth for the Higgs boson production in LHC.

\section{THE GAP SURVIVAL PROBABILITY}\label{sec:disc}

The main aspect discussed in this paper is a new mechanism to produce the SM Higgs boson in $pp$ collisions, which may increase the event rate in the case of nuclei collisions. The production cross section is enhanced by two or three orders for $pA$ collisions, showing its importance for the Higgs boson detection in LHC. However, the physics relative to diffractive processes was not totally discussed in this paper, since the GSP is still an open question in high-energy phenomenology, and certainly plays an important role in all predictions for diffractive processes in LHC. In the investigation of the Higgs boson production by DPE in UPC, in a more conservative approach, the GSP was introduced in the same way as employed by the Durham group, 3\%. Similarly, the group of Tel-Aviv has performed a more comprehensive calculation, obtaining a GSP of 0.4\% \cite{Miller:2006bi}, which is much smaller than that predicted by the Durham group. This result may change the scenario for the Higgs boson production in LHC, since it is possible that other mechanisms may overestimate the predictions for the cross section, as well as the event rate. The Tab. \ref{tab2} presents the different values of the GSP for the Higgs boson production in LHC implemented by each mechanism and its corresponding production cross sections. Clearly, independent of the approach to compute the GSP, the scenario is very competitive. Particularly, for the photoproduction mechanism, the GSP can be introduced in a more optimistic way, since the UPC has large impact parameters that may enhance the GSP. Then, based on previous evidences from the HERA data for central dijet events \cite{Kaidalov:2003xf}, we also make the calculation with a GSP of 10\%. Therefore, the photoproduction mechanism inspects the Higgs boson photoproduction in LHC with a GSP of 3\% and 10\% for both $pp$ and $pA$ collisions.

\begin{table}[t]
\centering
\begin{tabular}{||c|c|c||}\hline\hline
\textbf{Subprocess}              & GSP (\%)          & $\sigma_{pp}$ (fb) \\ \hline 
$I\hspace{-4pt}PI\hspace{-4pt}P$ & $\,\,$ 2.6 $\,\,$ & $\,\,$ 3.00 $\,\,$ \\ \hline
$I\hspace{-4pt}PI\hspace{-4pt}P$ & $\,\,$ 0.4 $\,\,$ & $\,\,$ 0.47 $\,\,$ \\ \hline
$\gamma\gamma$                   & $\,\,$ 100 $\,\,$ & $\,\,$ 0.12 $\,\,$ \\ \hline
$\gamma p$                       & $\,\,$ 3.0 $\,\,$ & $\,\,$ 1.77 $\,\,$ \\ \hline
$\gamma p$                       & $\,\,$ 10. $\,\,$ & $\,\,$ 5.92 $\,\,$ \\ \hline\hline
\end{tabular}
\caption{\label{tab2}Estimates for the GSP collected from different mechanisms for the Higgs boson production in LHC. The subprocess are identified as follows: $I\hspace{-4pt}PI\hspace{-4pt}P$ is the DPE approach carried out by the Durham group \cite{Forshaw:2005qp}, $\gamma\gamma$ is the two-photon process accounted by the group of Tel-Aviv \cite{Levin:2008gi}, and $\gamma p$ is the photoproduction mechanism \cite{GayDucati:2008zs}.}
\end{table}

\section{CONCLUSIONS}\label{sec:ccl}

In this work we present a new mechanism for the Higgs boson production at LHC. Similarly with the calculation performed by other approaches \cite{Forshaw:2005qp,Khoze:1997dr,d'Enterria:2009er}, the Higgs boson photoproduction is investigated in the $\gamma p$ subprocess in UPC, where the initial photon is emitted from the electromagnetic field of the colliding hadron, which can be a proton or a nucleus. As already done in other works, we include a survival probability of the rapidity gaps in the same way applied in the Pomeron-Pomeron mechanism, taking a value of 3\%. However, as we are interested in the $\gamma p$ subprocess, we inspect also the possibility for a higher GSP, like 10\%, which is based on the results of dijet events in HERA. Thus, we have found a production cross section of about 2 fb in $pp$ collisions for a GSP of 3\%, but this result is enhanced to about 6 fb for the higher GSP, higher to that predicted for the Pomeron-Pomeron mechanism. Moreover, extending this calculation to peripheral processes involving nuclei in LHC, we also have computed the cross section for $pA$ collisions. As a result, the production cross section is increased by a factor higher than 10$^{2}$ in comparison to $pp$ collisions, reaching about 0.6 pb for a Higgs boson mass of 120 GeV. In contrast to other results obtained for the Higgs boson production in $pA$ collisions, the photoproduction cross section is similar to that predicted by the two-photon mechanism, but higher than that obtained from the Pomeron-Pomeron process. Besides, we would like to comment that in comparison with the results from Ref.\cite{d'Enterria:2009er}, our results show a production cross section higher than the two-photon process in $pA$ collisions. This difference occurs due to the fact that the photoproduction approach has only one photon flux, which is less suppressed in high energies than the two-photon approach. Finally, to summarize our conclusions, we present in Tab. \ref{tab3} the expected number of events per year in this mechanism, considering the leading Branching Ratio BR($H \to b\bar{b}$) $\approx$ 72\% for a Higgs mass of 120 GeV \cite{Ahrens:2010rs}.

\begin{table}[t]
\centering
\begin{tabular}{||c|c|c|c|c||}\hline\hline
                & $\sigma$ (fb) & BR $\times \, \sigma$ (fb) & ${\cal{L}}$ (fb$^{-1}$) & events/year \\ \hline 
$pp$            & 1.77          & 1.27                       & 1(30)                   & 1 (30)      \\ \hline
$pp$            & 5.92          & 4.26                       & 1(30)                   & 6 (180)     \\ \hline
$pPb$           & 617           & 444                        & 0.035                   & 21          \\ \hline
$pPb$           & 2056          & 1480                       & 0.035                   & 72          \\ \hline\hline
\end{tabular}
\caption{\label{tab3}Expected number of events per year for the Higgs boson detection with $M_{H}$=120 GeV in LHC. The Branching Ratio (BR) of the Higgs boson decay into a $b\bar{b}$ pair is about 72\% \cite{Ahrens:2010rs}. The beam luminosity ${\cal{L}}$ are taken from Ref.\cite{D'Enterria:2007xr} for the proton beam, and from Ref.\cite{Carminati:2004fp} for the lead beam.}
\end{table}

As the GSP is still under study for the collisions in LHC, different approaches for this probability may increase the production cross section, and all results may become very competitive in the LHC kinematical regime. Therefore, the photoproduction mechanism becomes a very promising process to look for the Higgs boson in LHC.

\section{ACKNOWLEDGEMENTS}
MBGD and GGS would like to thank E. Levin, E. Gotsman, E.G.S. Luna and R. Sassot for helpful discussions. Moreover, we would like to thank D.~G. D'Enterria for valuable comments about this work. This work is partially supported by CNPq.

\end{document}